# Extraction Of Flat And Nested Data Records From Web Pages

P.S Hiremath[1], Siddu P. Algur[2]
1 Dept. of P.G. Studies and Research in Computer Science, Gulbarga University, Karnataka, India.
2 BVB College of Engg. and Tech., Department of Information Science and Engg., Hubli, Karnataka, India.
Email: hiremathps@hotmail.com, hiremathps53@yahoo.com algursp@bvb.edu,siddu_p_algur@hotmail.com.

**Abstract**

This paper studies the problem of identification and extraction of flat and nested data records from a given web page. With the explosive growth of information sources available on the World Wide Web, it has become increasingly difficult to identify the relevant pieces of information, since web pages are often cluttered with irrelevant content like advertisements, navigation-panels, copyright notices etc., surrounding the main content of the web page. Hence, it is useful to mine such data regions and data records in order to extract information from such web pages to provide value-added services. Currently available automatic techniques to mine data regions and data records from web pages are still unsatisfactory because of their poor performance. In this paper a novel method to identify and extract the flat and nested data records from the web pages automatically is proposed. It comprises of two steps : (1) Identification and Extraction of the data regions based on visual clues information. (2) Identification and extraction of flat and nested data records from the data region of a web page automatically. For step1, a novel and more effective method is proposed, which finds the data regions formed by all types of tags using visual clues. For step2, a more effective and efficient method namely, Visual Clue based Extraction of web Data (VCED), is proposed, which extracts each record from the data region and identifies it whether it is a flat or nested data record based on visual clue information – the area covered by and the number of data items present in each record. Our experimental results show that the proposed technique is effective and better than existing techniques.

## I. INTRODUCTION

These days most of the companies manage their business through web sites and use these web sites for advertising their products and services. These data which are dynamic need to be collected and organized such that after extracting information from these data one can produce many value-added applications. For example, in order to collate and compare the prices and features of products available from the various Web sites, we need tools to extract attribute descriptions of each product (called data object) within a specific region (called data region) in a web page. If one examines a web page (as illustrated in Fig. 1) there are many irrelevant components intertwined with the descriptions of data objects. These items include advertisement bar, product category, search panel, navigator bar, and copyright statement, etc. In many web pages, there are normally more than one data object intertwined together in a data region, which makes it difficult to discover the attributes for each page. Furthermore, since the raw source of the web page for depicting the objects is non-contiguous one, the problem becomes more difficult. In real applications, the users require the description of individual data object from complex web pages derived from the partitioning of data region. There are different approaches in practice due to Hammer, Garcia Molina, Cho, and Crespo [1], Kushmerick [2], Chang and Lui [3], Crescenzi, Mecca, and Merialdo [4], Zhao, Meng, Wu and Raghavan [5] which address the problems of web data extraction through wrapper generation techniques.

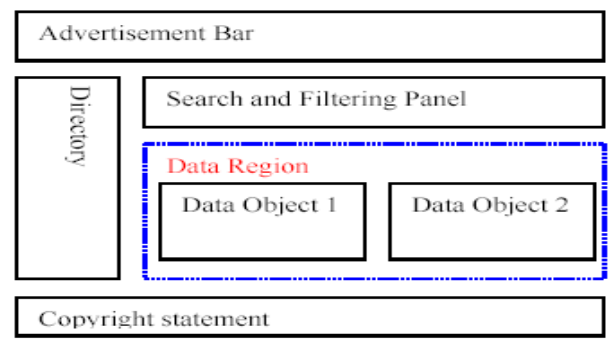

**Fig. 1 A schematic view of a webpage**

In the first approach due to Hammer, Garcia Molina, Cho, and Crespo [1], one has to manually write an extraction program for each web site based on observed format patterns of the site. As this is manual approach, it is very laborious, labor intensive and time consuming and does not scale to a large number of sites. The second approach due to Kushmerick [2] is wrapper induction or wrapper learning, and is the main technique in practice. In [], the user first manually labels a set of trained pages. A learning system then generates rules from the training pages. The resulting rules are then applied to extract target items from web pages. These methods either require prior syntactic knowledge or substantial manual efforts. An example of wrapper induction systems is WEIN by Baeza Yates [6]. The third approach due to Chang and Lui [3] is the automatic approach. Since structured data objects on the web are normally database records retrieved from underlying web databases and displayed in web pages with some fixed templates, this automatic method aims to find patterns/grammars from the web pages and then use them to extract data. Examples of automatic systems are – IEPAD [3] and ROADRUNNER [4]. The fourth approach is as applied for MDR by Liu, Grossman, and Zhai [7] which basically exploits the regularities in the HTML tag structure directly. It is often very difficult to derive accurate wrappers entirely based on HTML tags. The MDR algorithm makes use of the HTML tag tree of the web page to extract data records from the page. However, erroneous tags in the HTML source pages may result in building of incorrect trees, which in turn makes it impossible to extract data records correctly. DEPTA by Zhai, and Liu [8] uses visual information (locations on the screen at which the tags are rendered) to infer the structural relationship among tags and to construct a tag tree. But this method of constructing a tag tree has the limitation that, the tag tree can be built correctly only as far as the browser is able to render the page correctly. The computation time for constructing the tag tree is also an overhead. Further, this method also fails to identify some of the data records. NET by Bing Liu and Y. Zai [9] extracts data from web pages that contain a set of flat or nested data records automatically in two steps. This approach also depends on building of tag tree and post order traversal of the tag tree to identify data records at different levels.

To over come some of the problems in these methods discussed above, we propose a novel and more effective method to extract data regions from web pages that contain a set of flat or nested data records. This proposed technique is implemented in two steps :

i) Given a web page in the first step the identification and extraction of the data region based on visual clue (location of data region/ data records/ data items on the screen at which the tags are rendered) information of web pages is taken up using an algorithm here after called as VSAP (Visual Structure based Analysis of web Pages[12] ).

ii) Given a relevant data region (i. e., the output of VSAP algorithm) in the second step identification and extraction from flat and nested data records is carried out based on visual clues information using an algorithm, here after called as VCED (Visual Clue based Extraction of web Data[13] ).

The paper is organized into five sections. The section 2 presents the related work reported in the literature. The section 3 deals with the proposed technique. The section 4 gives the experimental results. Finally the conclusions are given in section 5.

## II. RELATED WORK

Present methodologies which are in place to deal with the problem of data extraction from web pages are presented in brief in this paragraph. For automatic extraction, from web pages Crescenzi, Mecca, and Merialdo [4], Zhao, Meng, Wu and Raghavan [5], Lerman, Getoor, Minton, and Knoblock [10], used to find patterns or grammars from multiple pages containing similar data records. They require an initial set of pages containing similar data records which is a limitation. Lerman, Getoor, Minton and Knoblock, [10], propose a method that tries to explore the detail information pages behind the current page to segment the data records. The need for such detail pages is a drawback because many data records do not have such pages. Chang and Lui [3], use string matching method, but had limitation of not finding nested data records. A similar methods are proposed by Wang, Lochovsky [11]. Liu, Grossman, and Zhai [7] and Zhao, Meng, Wu and Raghavan [5]. In some of these algorithms proposed uses the method of identification of data records, but these cannot extract data items from the data records and do not handle nested data records. DEPTA by Zhai and Liu [8] is able to align and extract data items from the data records but does not handle nested data records. The NET by Bing Liu and Y. Zai [9] which is the latest and widely used at present (Nested data Extraction using Tree matching) works in two main steps: (i) Building a tag tree of the page: Due to numerous tags and unbalanced tags in the HTML code of the page, building a correct tag tree is a complex task. A Visual based method is used to deal with this problem. (ii) Identifying data records and extracting data from them. The algorithm performs a post order traversal of tag tree to identify data records at different levels. This ensures that nested data records are found. The tree edit distance algorithm and visual clues are used to perform these tasks. Though this technique is able to extract the flat or nested data records, construction of tag tree and its post order traversal is consider to be an overhead.

All above automatic methods are tag dependant, incorporate and involve time-consuming tag tree construction. These are based on many assumptions which do not always hold good for all web pages and hence are not accurate. The proposed method is such that it does not make any such assumptions and can scale well for almost all web pages. It is also independent of the type of tags and dispenses with the time-consuming tag tree construction procedure.

## 3. THE PROPOSED TECHNIQUE

The system model for the proposed technique which is a combination of VSAP and VCED is shown in Fig. 2

It consists of the following components.

VSAP :

Determination of bounding rectangles

Identification of the data regions

Largest rectangle identifier

Container identifier

Data region identifier (Filter)

VCED :

Data record extractor.

Data record identifier – flat or nested.

The HTML page is input to the VSAP sub system, which extracts the relevant data regions and these data regions are input to the VCED subsystem which identifies & extracts the flat and nested data records. The output of each component is the input for the next component. This section focuses on the first step, identification and extraction of data regions.

## 3.1. Data Region Extraction

The algorithm used here is called VSAP (Visual Structure based Analysis of web Pages). The visual information (i.e, the locations on the screen at which tags are rendered) helps the system in three ways:
(a) It enables the system to identify gaps that separate data records, which helps to segment data records correctly, because the gap within a data record (if any) is typically smaller than that in between data records. (b) The visual or display information also contains information about the hierarchical structure of the tags. (c) By the visual structure analysis of the web pages, it can be analyzed that the relevant data region seems to occupy the major central portion of the web page.

The VSAP technique is based on three *observations*:

(a) A group of data records, that contains descriptions of a set of similar objects, is typically presented in a contiguous region of a page. (b) The area covered by a rectangle that bounds the data region is more than the area covered by rectangles bounding other regions, eg., advertisements and links. (c) The height of a irrelevant data record within a collection of data records is less than the average height of relevant data records within that region.

**Fig. 2 Proposed system model.**

The above observations are inconsistent with experimental results with the definition of data region and data record as defined below.
A Data region is defined as the most relevant portion of a web page.
e.g., A region on the product-related web-site that contains a list of products forms the *data region.*
A Data record is defined as a collection of data that together represents a meaningful independent entity.
eg., A product listed inside a data region on a product-related website is a data record
Fig 3 illustrates an example, which is a segment of a web page that shows a data region containing list of four books. The full description of each book is a data record.

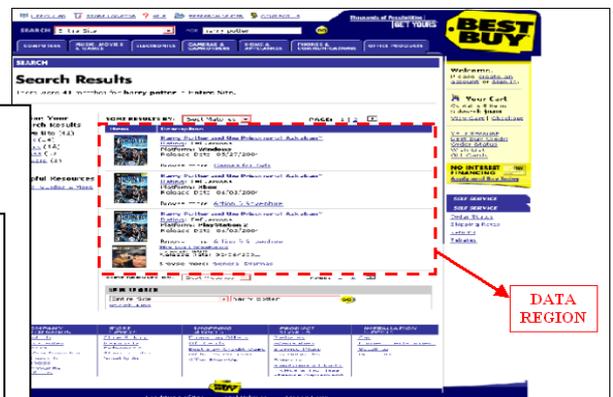

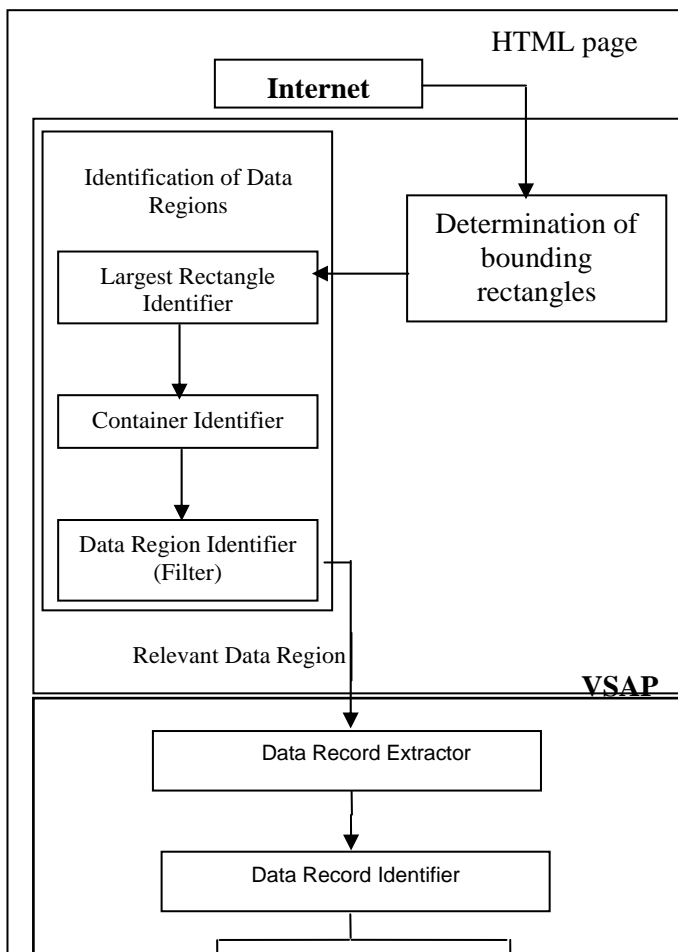

**Fig. 3 An example of a Data Region containing 4 data records**

With this definitions of data regions and data record, the VSAP algorithm for the proposed technique is :

Algorithm VSAP(*HTML document*)

1. Set maxRect=NULL
2. Set dataRegion=NULL
3. FindMaxRect(*BODY*);
4. FindDataRegion(*maxRect*);
5. FilterDataRegion(*dataRegion*);

The lines 1 and 2 specify initializations. The line 3 finds the largest rectangle within the container. Line 4 identifies the data region which consists of the relevant data region and some irrelevant regions also. Line 5 identifies the actual relevant data region by filtering the bounding irrelevant regions.

The two main steps of VSAP algorithm, namely, i) Determination of bounding rectangles and

ii) Identification of data regions are explained below.

### 3. 1.1. Determination of bounding rectangles

In the first step of the proposed technique, the co-ordinates of all the bounding rectangles in the web page are determined. The VSAP approach uses the MSHTML parsing and rendering engine of Microsoft Internet Explorer 6.0. This parsing and rendering engine of the web browser gives co-ordinates of a bounding rectangle. The MSHTML parsing and rendering engine is the main HTML component of the Microsoft Internet Explorer web browser. The IE rendering engine works as a COM component. This rendering engine of the browser produces the boundary coordinates.

HTML files are scanned for tags. For each tag encountered, the co-ordinate of the top-left corner, height and width of the *bounding rectangle* of that tag are determined inline with, definition. Every HTML tag specifies a method for rendering the information contained within it. For each tag, there exists an associated rectangular area on the screen. Any information contained within this rectangular area obeys the rendering rules associated with the tag. This bounding rectangle is constructed by obtaining the co-ordinate of the top-left corner of the tag, the height and the width of that tag. The left and top co-ordinates of the tag are obtained from the offsetLeft and offsetTop properties of the HTMLObjectElement. These values are with respect to its parent tag. The height and width of that tag are available from the offsetHeight and offsetWidth properties of the HTMLObjectElement Class.

Fig. 4 shows a sample web page of a product-related website, which contains the list of books and their descriptions which form the data records inside the data region.

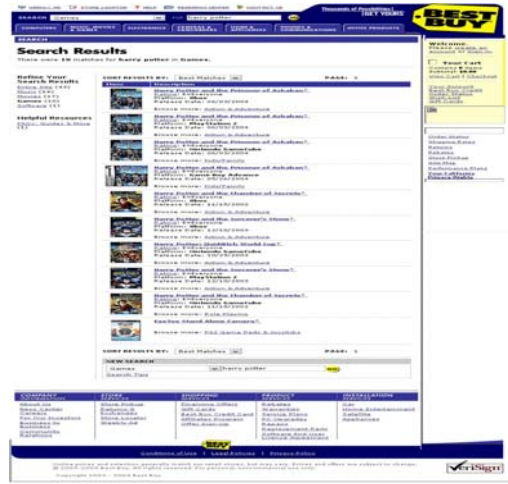

**Fig. 4 A Sample Web Page of a product related web-site**

For each HTML tag on the web page, there exists an associated rectangular area on the screen, which forms the bounding rectangle for that specific tag. Fig 5 shows the bounding rectangles for the <TD> tags of the web page shown in Fig 4.

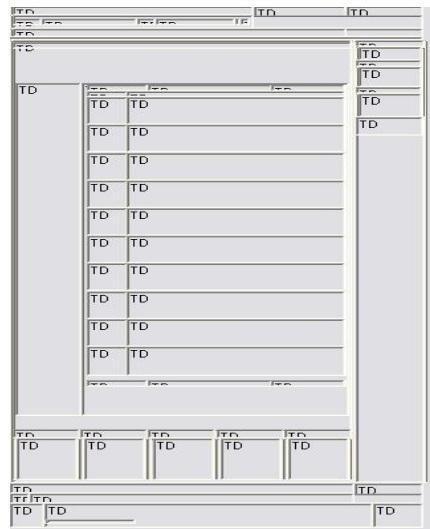

**Fig. 5 Bounding rectangles for <TD> tag corresponding to the web page in Fig. 3**

Figure 6 shows the top-left corner's co-ordinate, height and width obtained for constructing the bounding rectangle of a <TD> tag. The bounding rectangle of that tag is shaded. The co-ordinates (80, 45), specify the corresponding height and width of the bounding rectangle of the <TD> tag, respectively.

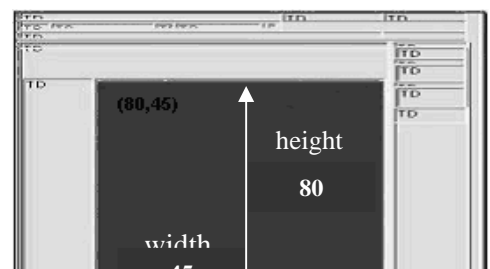

39

**Fig. 6 Bounding rectangle of <TD> tag**

### 3.1.2. Identification of the data regions

The second step of the proposed technique is to identify the data region of the web page. The data region is the most relevant portion of a web page that contains a list of data records.

The three steps involved in identifying the data region are:

Step i) Identify the largest rectangle.

Step ii) Identify the container within the largest rectangle.

Step iii) Identify the data region containing the data records within this container

These steps are explained below

(i) Identification of the largest rectangle:

Based on the height and width of bounding rectangles obtained in the previous step, determine the area of the bounding rectangles of each of the children of the BODY tag. Then determine the largest rectangle amongst these bounding rectangles, the reason being the largest bounding rectangle will always contain the most relevant data in that web page. Thus, by determining the largest rectangle, superset of the data region can be obtained. In Fig 7 the example of largest rectangle is shown with a dotted border.

The procedure *FindMaxRect* identifies the largest rectangle amongst all the bounding rectangles of the children of the BODY tag, as follows.

Procedure FindMaxRect (*BODY*)

   for each child of BODY tag

 Begin

  find the co-ordinates of the bounding rectangle for the child

   if the area of the bounding rectangle > area of maxRect then

          maxRect = child

   endif

 end

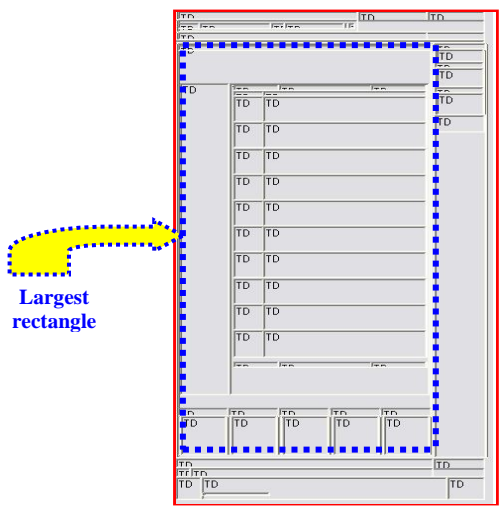

**Largest rectangle**

Fig. 7 Largest Rectangle amongst bounding rectangles of Children of BODY tag

(ii) Identification of the container within the largest rectangle:

Once the largest rectangle are obtained a set of all the bounding rectangles whose area is more than half the area of the largest rectangle are formed. The rationale behind this is that the most important data of a web page must occupy a significant portion of the web page. Next, determine the bounding rectangle having the smallest area in this set. The reason for determining the smallest rectangle within this set is that the smallest rectangle will only contain data records. Thus a *container* (A *container* is a superset of the data region which may or may not contain irrelevant data) is obtained. It contains the data region and some irrelevant data. For example, the irrelevant data contained in the container may include advertisements on the right and bottom of the page and the links on the left side. The Fig 8 shows an example of container identified from the web page shown in Fig 4.

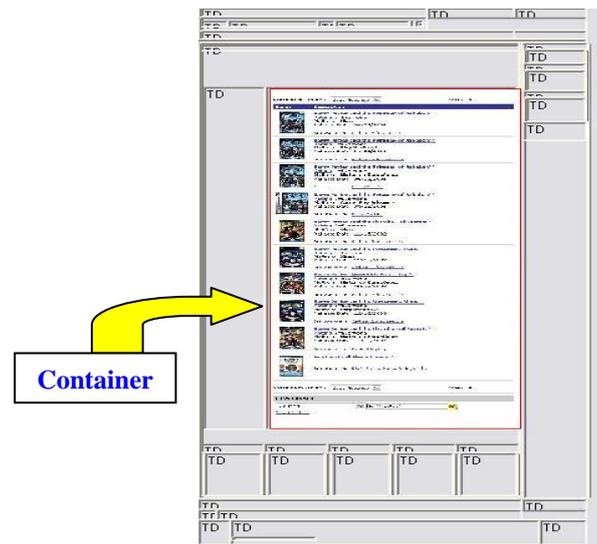

**Container**

**Fig. 8 The container identified from sample web page in Fig. 4**

The procedure *FindDataRegion* identifies the container in the web page which contains the relevant data region along with some irrelevant data also, as shown below :

Procedure FindDataRegion(*maxRect*)
ListChildren=depth first listing of the children of the tag associated with maxRect

   for each tag in ListChildren

 Begin

  If area of bounding rectangle of tag > half the area of maxRect then

   If area of bounding rectangle dataRegion > area of bounding

     rectangle of tag then

      dataRegion = tag

```
            endif
        endif
    end
```

The Fig 9 shows the enlarged view of the container shown in the Fig 8. We note that there is some irrelevant data, both on the top as well as the bottom of the actual data region containing the data records.

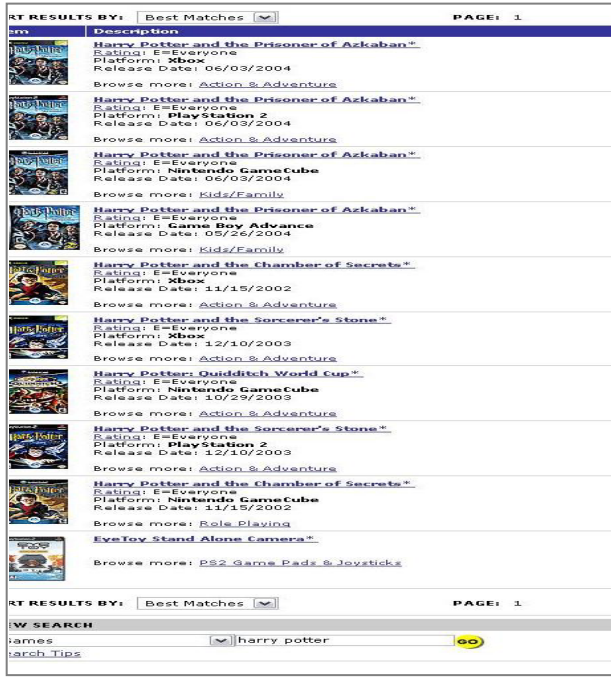

**Fig. 9 The Enlarged view of the Container shown in Fig. 8**

(iii) Identification of data region containing data records within the container:

To filter the irrelevant data from the container a *filter* is used. The filter determines the average heights of children within the container. Those children whose heights are less than the average height are identified as irrelevant data and are filtered off. The Fig 10 shows a filter applied on the container in Fig 9, in order to obtain the data region. We note that the irrelevant data in this case is on the top and bottom of the container, which are being removed by the filter.

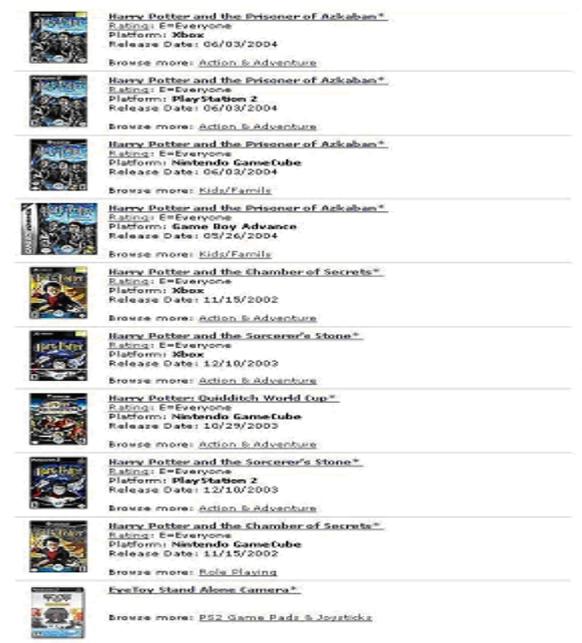

**Fig. 10 Data Region obtained after filtering the Container in Fig. 9**

The procedure FilterDataRegion filters the irrelevant data from the container, and gives the actual data region as the output. It is as follows:

Procedure FilterDataRegion (dataRegion)

    totalHeight = 0

    for each child of dataRegion

      totalHeight += height of the bounding rectangle of child

    end

    avgHeight = totalHeight / no of children of dataRegion

    for each child of dataRegion

    Begin

    If height of child's bounding rectangle < avgHeight then

        Remove child from dataRegion

    endif

    end

The VSAP technique, as described above, is able to mine the relevant data region containing data records from the given web page efficiently. The extracted relevant data region will become an input to VCED algorithm, the proposed technique for extraction of flat and nested data records.

### 3.2. Data Record Extraction

In this section a more effective method to identify and extract flat and nested data records from a given web page automatically is proposed. The method is called VCED. The Fig.11 illustrates an example, which is a segment of a web page that shows flat and nested data records.

By Definition flat and nested data records are defined as below :

A *flat data* record is defined as a collection of data items that together represents a single meaningful entity. eg., the product having single size, look, price etc.,

A *nested data record* is defined as one that provides multiple description of the same entity. Eg., the same type of products but different sizes, looks, prices etc.,

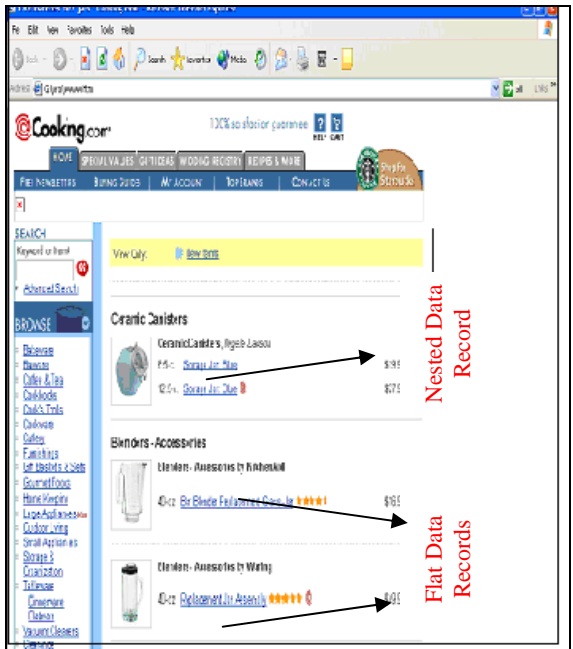

**Fig. 11 An example of flat and nested data records**

VCED algorithm is described by considering a sample webpage as shown in Fig. 11. When a web page having description of products is given to VSAP, it identifies and extracts the data region. All the noises of a given web page are eliminated using filter. The filtered data region corresponding to the Fig. 11. is shown in Fig.12.

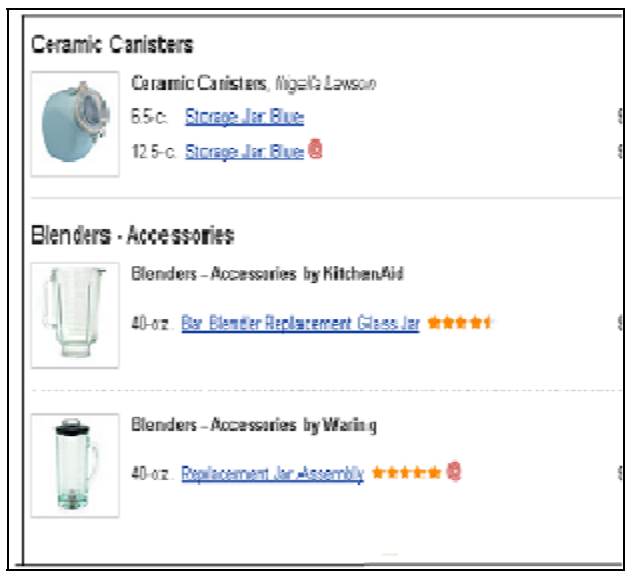

**Fig. 12 Filtered data region**

The filtered data region is given as the input to VCED system which extracts the flat and nested data records from the given data region and extracts data fields from the identified records.

### 3.2.1. Data record Extractor

Extraction of data records is based on visual clues. In the first step of the proposed technique, we determine the height of all the data records. This approach uses the MSHTML parsing and rendering engine that gives the height of each data record. The height of the data record is obtained from the offsetHeight property of the HTMLObjectElement. Next, the average height of the records is calculated. The average height of all the records provides the approximate height of each record. The height of each data record is compared with the average height. If the height of the child is greater than or equal to the average height, then the data record is extracted.

The procedure Extract Data Record extracts the flat and nested records from given data region. It is as follows.

Procedure ExtractDataRecord(dataRegion)
{
   THeight=0
     For each child of dataRegion
      BEGIN
        THeight += height of the bounding rectangle of child
      END
      AHeight = THeight/no of children of dataRegion
     For each child of dataRegion
    BEGIN
     If height of child's bounding rectangle > AHeight
     BEGIN
      dataRecord=child
     END
    END
}

The Fig. 13 shows extracted data records from the data region shown in the Fig.12.

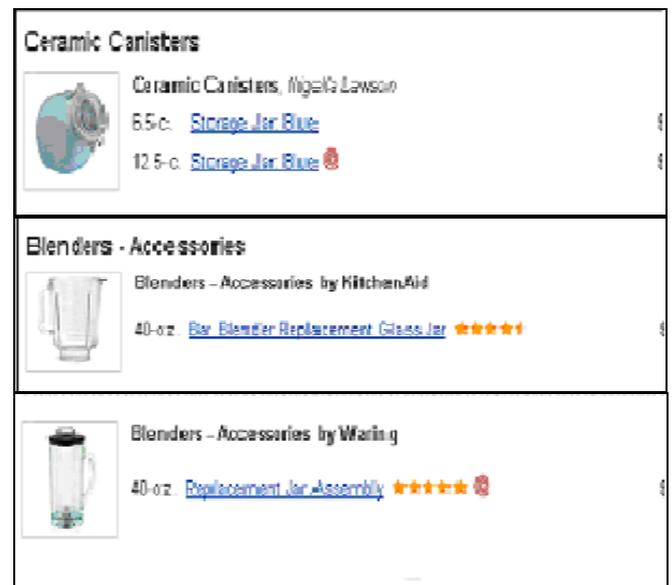

**Fig. 13 Extracted data records**

### 3.2.2. Data record Identifier

Identification of data records, as flat or nested, is essential in order to simplify the task of extracting the data items, which is very useful for various applications as mentioned earlier.

This technique determines the data fields for each data record within the data region. Various tags such as <TD>, <TR>, <A>, , represent the data fields. By counting these tags as they are encountered, the number of fields is obtained. The flat record gives description of a single entity, whereas the nested data record gives multiple description of a single entity, so the data fields in flat records are less as compared to that of nested records. Experimental observations have shown that the number of fields in the nested data records is at least 40% (approx) more than that of the flat records. The number of fields in the first record is compared with the number of fields in the next record. If the number of fields is more than 40%, then it is a nested record else it is a flat record. Suppose a condition is encountered where the number of fields is equal then in both cases. Then the record is compared with the third record and so on until the condition is satisfied.

The procedure IdentifyNestedData identifies whether the record is flat or nested based on the number of data items present in the data record. It is as follows:

Procedure IdentifyNestedData(dataRecord[I], dataRecord[I+1])

{

  noofField[I]=0

   For I 1 to no of records

    BEGIN

     noofFields [I]= noofFields[I]+noofFields in the record[I]

    END

  DO

   For I 1 to no of records

    BEGIN

     For dataRecord [I], dataRecord[I+1]

      IF the no of fields in the [I+1]$^{th}$ record>=40% of the no of fields in the [I]$^{th}$ record

       Then

        [I+1]$^{th}$ record is a  nested data record

      ELSE

       The [I]$^{th}$ record is a nested  data record

    END

  WHILE (EOF)

}

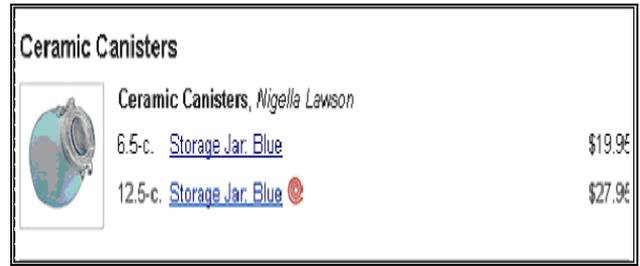

**Fig. 14(a) Identified nested data record, No. of data fields=12**

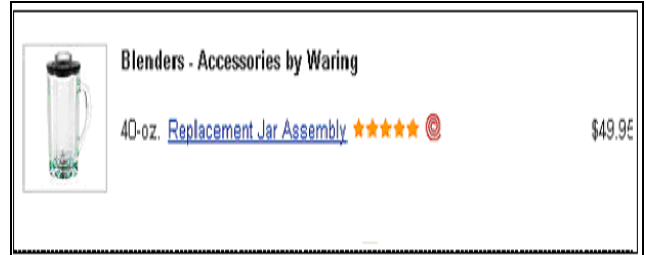

**Fig. 14(b) Identified flat data record, No. of data fields = 7**

The Fig. 14 shows the identified nested and flat data records. In Fig.14 (a), the number of data fields is 12 and in Fig.14 (b) the number of data fields is 7. The number of data fields in Fig 14 (a) is 58.3% more than the number of data fields in Fig 14 (b).

### 4. EXPERIMENTAL RESULTS

In this section, we compare the proposed technique with two state-of-the-art existing systems NET [9] (which is an improvement of MDR [7] . We do not compare it with the method in [3] and method in [4] here as it is shown in [7] that MDR is already more effective than them.

In running MDR, their default settings were used. MDR system was downloaded at *http://www.cs.uic.edu/~liub/MDR/ MDR-download.htm.* The *recall* and *precision* measures (which are widely used to evaluate information retrieval systems) to evaluate the performance of VSAP system for extracting data records were used. Recall and precision are defined below:

*Recall* = Ec / Nt  and  *Precision* = Ec / Et

where, Ec is the total number of correctly extracted records , Nt is the total number of records on the page , and Et is the total number of records extracted. *Recall* defines the correctness of the data records identified and *precision* is the percentage of the relevant data records identified from the web page.

The experimental results are given in Table 1. Correct results (Cor.) are relevant data records present on the page that have been correctly identified. Wrong results (Wr.) are irrelevant data records that have been incorrectly identified. The results obtained after running both MDR and VSAP are shown in table 1. The successes of both algorithms are compared in terms of precision and recall.

In table 1 the column 1 shows the lists the URL of each site. In some sites more than one page are tried (which have different data record formats). All our experimental web pages are selected randomly. Due to long URLs of some pages it was not possible to completely present them here. *Columns 2 and 4* give the numbers of correct (Cor.) records extracted by MDR and VSAP from the pages of each site respectively. These data records are the relevant ones (e.g., product lists). They do not include navigation areas, advertisements etc, which may also have regular patterns. MDR cannot handle nested data records (records within records), but VSAP is able to handle such data records also. *Columns 3 and 5* give the numbers of data records extracted wrongly (Wr.) by MDR and VSAP, from the pages of each site respectively. *x/y* means that *x* is the number of extracted results that are wrong, and *y* is the number of results that are not extracted.

| URL | MDR | | VSAP | |
|---|---|---|---|---|
| | Cor. | Wr. | Cor. | Wr. |
| http://www.tigerdirect.com/…. | 8 | 36/ 0 | 8 | 0/0 |
| http://www.amazon.com/…. | 0 | 17/25 | 25 | 1/0 |
| http://www.cooking.com/…. | 17 | 13/3 | 20 | 0/0 |
| http://www.ebay.com/….. | 25 | 30/0 | 25 | 0/0 |
| http://www.powells.com/….. | 9 | 47/1 | 10 | 5/0 |
| http://www.barnesandnoble.com/…. | 10 | 30/0 | 10 | 0/0 |
| http://www.pricegrabber.com/….. | 0 | 0/25 | 25 | 0/0 |
| http://www.shoebuy.com/…. | 12 | 12/84 | 96 | 0/0 |
| http://www.smartbuy.com/….. | 0 | 15/10 | 10 | 0/0 |
| http://www.reviews.cnet.com/….. | 24 | 38/1 | 25 | 3/0 |
| http://www.nothingbutsoftware.com/ | 10 | 12/0 | 10 | 0/0 |
| http://www.refurbdepot.com/…. | 0 | 6/15 | 15 | 0/0 |
| http://www.drugstore.com/…. | 15 | 14/0 | 15 | 0/0 |
| http://www.bookpool.com/…. | 10 | 7/0 | 10 | 0/0 |
| http://www.target.com/…… | 0 | 0/12 | 12 | 1/0 |
| Total | 140 | 277 / 176 | 316 | 10 / 0 |
| 1.1.1 Recall | 33.5% | | 96.93% | |
| 1.1.2 Precision | 44.3% | | 100% | |

**Table 1. Comparison of VSAP and MDR**

The results shown are obtained without using any form of content search. All the URLs given are obtained from the search functionality in the browser, where the search query for all URLs is "watch" (except for test case 12 where "camera" was specified, test case 14 where "computer" was specified and 8 where "men's" was specified)

It can be observed that test cases 7 and 15 hanged in case of MDR and hence have been considered as missing the correct data records. It has been experimentally found out that MDR fails to identify a single data record where as VSAP successfully identifies it for the URL,

*http://www.nothingbutsoftware.com/SiteSearch.asp?db=12719&query=watches*

The last three rows of Table 1 give the total of each column, the recall and precision of each system. For MDR and VSAP, the recall and precision are computed based on the total number of data records found in all pages and the actual number of data records in these pages. It can be seen that the data record extraction is highly effective. It is also observed that VSAP performs significantly better than MDR.

The results obtained after running VCED are presented in Table 2 for few URL's. These are compared with the state-of-the-art existing system NET by Bing and Yanhong [9]. The results are not compared with DEPTA by Zhai and Liu [8] here as it is shown that NET is better than DEPTA. For flat and nested data records, the proposed method performs very well.

In table 2 Column 1, lists the site of each test page. Due to the space limitations, all the URL's considered for experimentation are not listed. In working the authors have only considered non erroneous pages for testing because such pages are relatively rare and quite difficult to find. Column 2 and 4 give the number of data items extracted wrongly (Wr) by NET and proposed method from each page respectively. In x/y, x is the number of extracted results that are incorrect and y is the number of results that are not extracted. Columns 3 and 5 give the numbers of correct (Corr) data items extracted by NET and proposed method from each page respectively. Here, in x/y, x is the number of correct items extracted and y is number of items in the page. From the table, it is observed that, for flat and nested data records, the proposed method performs better than the other. The precision and recalls are computed based on extraction performed on all test pages.

| URL | NET | | VCED | |
|---|---|---|---|---|
| | Wr. | Corr. | Wr. | Corr. |
| Without Nesting | | | | |
| http://www.bookpool.com | 0/0 | 15/15 | 0/0 | 15/15 |
| http://www.amazon.com | 0/0 | 22/22 | 0/0 | 22/22 |
| http://www.shopping.com | 0/0 | 20/20 | 0/0 | 20/20 |
| http://www.barnesandnobles.com | 0/0 | 10/10 | 0/0 | 10/10 |
| http://www.cooking.com | 0/0 | 28/29 | 0/0 | 29/29 |
| http://tigerdirect.com | 0/0 | 12/14 | 0/0 | 13/14 |
| http://www.kmart.com | 0/0 | 70/70 | 0/0 | 70/70 |
| Recall | | 97.15% | | 98.99% |
| Precision | | 99.3% | | 98.92% |
| With Nesting | | | | |
| http://www.amazon.com | 0/0 | 22/25 | 0/0 | 25/25 |
| http://www.kmart.com | 1/0 | 42/43 | 0/0 | 43/43 |
| http://www.cooking.com | 1/0 | 62/63 | 0/0 | 62/63 |
| Recall | | 98.63% | | 100% |
| Precision | | 99% | | 100% |

**Table 2. Comparison of VCED and NET**

## 5. CONCLUSION

In this paper, a technique is proposed which is more effective one to perform the automatic extraction structured data from web pages. Although the problem has been studied by several researchers, existing techniques make many strong assumptions or not effective. A novel and effective method based on a combination of VSAP and VCED is proposed to mine the data region and data items from the flat and nested data records in a web page automatically. It is a pure visual structure oriented method that can correctly identify the data region and data items from flat and nested data records. The proposed technique exploits the merits of both VSAP and VCED leading to better performance in data extraction from a webpage. The experimental results demonstrate the effectiveness of the proposed method.

.